%% file: TFPIE13_Caldwell_proceedings.tex
\documentclass[submission,copyright,creativecommons]{eptcs}
\providecommand{\event}{TFPIE 2013} % Name of the event you are submitting to

\include{header}

\usepackage{mathtools}

\date{\today}
\title{Structural Induction Principles for Functional Programmers} 
\author{James Caldwell\institute{Department of Computer Science\\University of Wyoming\\ Laramie, WY 82071}}
\def\titlerunning{Structural Induction Principles for Functional Programmers}
\def\authorrunning{James Caldwell}

\begin{document}
\maketitle
\begin{abstract}
  User defined recursive types are a fundamental feature of modern
  functional programming languages like Haskell, Clean, and the ML
  family of languages.  Properties of programs defined by recursion on
  the structure of recursive types are generally proved by structural
  induction on the type. It is well known in the theorem proving
  community how to generate structural induction principles from
  data type declarations. These methods deserve to be better know in
  the functional programming community. Existing functional
  programming textbooks gloss over this material. And yet, if
  functional programmers do not know how to write down the structural
  induction principle for a new type - how are they supposed to reason
  about it?  In this paper we describe an algorithm to generate
  structural induction principles from data type declarations. We also
  discuss how these methods are taught in the functional programming
  course at the University of Wyoming.  A Haskell implementation of
  the algorithm is included in an appendix.

\end{abstract}

\section{Introduction}

A fundamental claim made for functional programs is that they are easier to
reason about.  This is largely true because:
\begin{description}
\item{i.)} the evaluation mechanism is substitution based (following ordinary
  mathematical practice), and
\item{ii.)}  structural induction provides a straightforward mechanism
  for reasoning about programs defined by recursion on algebraic data
  types.
\end{description}
A recursive type definition naturally gives rise to a structural
induction principle for the type. For functions defined by recursion
on the structure of a type, structural induction is the natural
mechanism for reasoning about those functions.  A functional
programming course is the obvious place to make the relationship
between induction and recursion explicit, and yet, we know of no
standard text suitable for undergraduates that does so.  In fact,
there is no other point in the undergraduate curriculum where the
concrete relationship between induction and recursion can be made as
explicit as it can be in a course on functional programming.

In this paper we introduce the functional programming course as taught
at the University of Wyoming, briefly discuss well-founded induction,
the justification for structural induction, and then describe an
algorithm for generating a structural induction schema from a data
type declaration.  We also discuss the pedagogical approach we use at
the University of Wyoming to teach this material and give a few
examples.

\section{Functional Programming in Wyoming}

The University of Wyoming offers an ABET \footnote{ABET is the recognized
  accreditation board for college and university programs in applied science, computing,
  engineering, and technology programs in the U.S.}  is accredited Bachelor of
Science in Computer Science.  Functional Programming (COSC 3015) is a required
third year course for undergraduate students in the Computer Science degree.
% This is a somewhat unusual requirement for an undergraduate Computer Science
% program in the U.S., 
The University of Wyoming has had the requirement for at
least fifteen years.  Functional Programming is taught once a year and is a
prerequisite for the senior level Principles of Programming Languages (COSC
4870).

The functional programming course has been Haskell based since at least 2006;
in earlier carnations it was Scheme and LISP based. Students who take the
course have already taken Discrete Structures (COSC 2300) which, as taught in
Wyoming, emphasizes mathematical proofs.  Students taking functional
programming will (in theory) already know predicate logic, and will have done
proofs by mathematical induction and possibly by complete induction.  There is
often a gap between the time a student takes Discrete Structure and Functional
Programming and so proof methods are reviewed for the first week or two of the
functional programming course. This is done by reviewing the the mathematical
definition of a function and extensional equality for functions. This provides
a nice segue into higher order functions and students apply the definitions to
reason about {\em{curry}} and {\em{uncurry}}.

Though the course does not follow any one text, over the years it has
been taught with the assistance of a variety of texts
\cite{Bird98,Thompson99,Hudak00,Hutton07,Lipovaca11}.  Bird's text
\cite{Bird98} was the first one used to teach the course when it
transitioned to Haskell and is still the book that is philosophically
closest in spirit to the course as taught today.  An objective for the
functional programming course is for students to learn to do proofs in
concert with program development\footnote{Students undoubtedly do not
  carry this practice with them to program developments beyond the
  functional programming course, but they learn a disciplined and
  formal way to think about and reason about programs.}  Of the texts
cited above, only Bird's book carries through the proof theme from the
beginning of the book to the end. Bird introduces a number of
induction principles and, by example, expects students to be able to
derive new structural induction principles from type declarations
\cite[Exercise 6.2.1, pp. 191]{Bird98}.  In the functional programming
class at Wyoming we make the relationship more explicit and expect
students to be able to write down the structural induction principle
for an arbitrary Haskell type and to use it to prove some simple
properties about programs defined by recursion on the type.

Among the other texts references in the course, Thompson's book
\cite[pp. 141]{Thompson99} introduces structural induction for finite lists and
touches on induction for infinite lists but does not discuss induction
principles for other structures. Thompson's first introduction is well after
lists have first been introduced.  Similarly, Hudak's book introduces list
induction \cite[pp. 131]{Hudak00} and natural number induction
\cite[pp. 141]{Hudak00}, other forms of induction are not discussed.  Hutton's
text relegates reasoning about programs to the last chapter
\cite[Chap.13]{Hutton07} where he introduces both natural number induction and
list induction.  Lipovaca \cite{Lipovaca11} never mentions induction at all.
In \cite{Felleisen_design}, Felleisen, Findler, Flatt and Krishnamurthi do not
discuss induction but the design recipe connecting the shape of the input data
with the shape of the program is a closely related topic.  All texts mentioned
have their strengths; but regarding the goal of teaching programming together
with the methods for reasoning about programs, Bird's book does it best.

Among the texts that have not been used in the course (though some have been on
the recommended readings list) Reade's book \cite[pp.185]{Reade89} contains the
only elementary explanation we know of regarding how structural induction
principles can be derived from the declaration of a new type. Like Bird,
Cousineau and Mauny \cite{Cousineau95} present a series of examples which serve
to indicate how the induction principles can be gleaned from the type
declaration.

Of course more advanced programming language texts explain these
topics in some detail \cite{Winskel93,Mitchell96,Pierce02}.

The derivation of structural induction principles has been best covered in the
theorem proving literature. The classic paper is Burstall's
\cite{Burstall69}. Paulson covers it in detail \cite[pp. 77-135]{paulson90}.
The implementations in Isabelle HOL prover \cite{Nipkow02} is excellent, as are
the accounts of structural induction principles automatically generated by the
Coq prover \cite{Bertot04,Chlipala13}.

Perhaps the point we'd most like to make here is that, if programming and
proving are to go hand-in-hand, understanding how to generate structural
induction principles from data-type declarations is essential knowledge.  It is
not hard to do nor is it hard to teach, and students who learn this are not
hamstrung when it comes to proving things about a new user defined type.

\section{Background}

\subsection{Recursive Type Declarations}

Modern functional programming languages all support some convenient form
declaring new recursive types.  In Haskell \cite[Section 4.2.1]{Haskell98} user defined
recursive data types are given by {\em{Algebraic Data Type Declarations}}.

The declarations of new recursive types are given by specifying, in some way,
constructor names and their signatures.  In dialects of ML and in Haskell, such
type declarations may also be polymorphically parameterized and appear as
follows.

\[ data\; T\; [tVars] = C_0\,[typeNames] \mid C_1\,[typeNames] \mid \cdots \mid C_m\,[typeNames] \]
\noindent{}where $T$ is the name of the new type being defined. The type name
$T$ is followed by a list $[tVars]$ of polymorphic type variable names, say
$[V_1,\cdots,V_k]$. If the type is not parameterized, then the list is
empty. The length of the list is the arity of $T$.  On the right side of the
declaration there are $m$ constructor names $C_i, 0\le{}i<m, m>0$.  Each
constructor is followed by a list $[typeNames]$ of the names of the types of
its parameters. The list may contain previously declared types, polymorphic
type variables, and completely parameterized recursive instances of the type
$T$.  The {\em{arity}} of the constructor $C_i$ is the length of the specified
parameter list.  A constructor name with no parameters is a constant of type
$T$.

Recursive type declarations of this kind can be interpreted as many-sorted
$\Sigma$-algebras and the finite instances of an inductively defined data type
are denoted by the freely generated terms of the algebra \cite{Mitchell96,Gallier85}.

If a data type $T$ has arity $0$ then it denotes a type of kind $*$. If
$T$ has arity $1$ then it is a type constructor and has kind
$*\rightarrow{}*$. If $T$ has arity $k$ it is a type constructor of
$k$ arguments and has kind
$\overbrace{*\rightarrow{}\cdots\rightarrow{}*}^{k \,
  {\mathit{arrows}}}$.  Note that by convention the function type
constructor ``$\rightarrow$'' associates to the right so
$*\rightarrow{}*\rightarrow{}*$ means
$*\rightarrow{}(*\rightarrow{}*)$.  If $S$ is a type (possibly
parameterized) of the form $(T \;T_1\;\cdots{}T_k)$ we write
$TyCon(S)$ to denote the type constructor used to create $S$, in this
case $T$.  If $S$ is a constant (not parameterized) then $TyCon(S) =
S$.

Note that the constructors provide the only means for building
instances of a recursive data-type of the kind described here.  The
implication is that every instance of the type arises from an
application of a constructor $C_i$ to appropriately typed arguments.

\subsection{Well-founded Induction}

Well-founded induction is a powerful and flexible form of
induction. It is based on an ordering given by a well-founded relation
over a type $T$.  To show that a recursive function defined over a
recursive type $T$ terminates, it is enough to show that there is some
well-founded relation $\prec$ for which every recursive call is on a
smaller instance of $T$ with respect to $\prec$.  Excellent accounts
of well-founded induction can be found in \cite{Winskel93,Mitchell96}.

\begin{definition}[Well-founded Relation]
  A binary relation on a set $A$ is {\em{well-founded}} iff there are no
  infinitely descending chains $\cdots \prec a_i \prec \cdots \prec a_1 \prec
  a_0$, {\em{i.e.}} there is no function $a:\nat\rightarrow{}A$ such that for
  all $i\in\nat$, $a(i+1)\prec{}a(i)$.
\end{definition}

Note that a well-founded relation need not be transitive \cite{Mitchell96}.
Thus, for example, the relation $i\prec{}j \definedAs{} j = i + 1$ is
well-founded but not transitive: $(1\prec{}2)$ and $(2\prec{}3)$ but
$(1\not\prec{}3)$.

\begin{theorem}[Well-founded Induction]
  Let $\prec$ be a well-founded binary relation on a set $A$ and let $P$ be a
  property of $A$, then
\[(\forall{}x\!:\!A.\, (\forall{}y\!:\!A.\; y\prec{}x \Rightarrow P(y))\Rightarrow  P(x))\Rightarrow \forall{}x\!:\!A.\,P(x)\]
\end{theorem}

If $\Gamma$ is a context, well-founded induction can be written in the form of
a derived proof rule as follows:
\begin{center}
\newSequentRule{\Gamma,\,x\!:\!A,\,\forall{}y\!:\!A.\; y\prec{}x \Rightarrow P(y) \vdash{}P(x) }{\Gamma\vdash\forall{}x\!:\!A.\,P(x)}
\end{center}

Pedagogically, well-founded induction is a bit more difficult to justify than
structural induction which is more concrete.

\subsection{Structural Induction for $\nat$}

To show that a property $P$ holds for all natural numbers, mathematical
induction often suffices.  This principle is presented as follows.
\[(P(0)\wedge{}\forall{}k:\nat. P(k)\Rightarrow{}P(k+1)) \Rightarrow
\forall{}j:\nat. P(j)\] As we shall see, mathematical induction is just an
instance of structural induction on the natural numbers.  

To see this, consider the following data type having two constructors, a
constant $Z:N$ and the successor function $S:N\rightarrow{}N$.
\begin{program*}
\>              data Nat = Z $\mid$ S Nat \\
\end{program*}
Replacing $Z$ for $0$ and $S$ for $(+1)$ yields the following:
\[(P(Z)\wedge{}\forall{}k:\nat. P(k)\Rightarrow{}P(S\,k)) \Rightarrow \forall{}j:\nat. P(j)\] 
This is simply the structural induction principle for $\nat$. 
As a proof rule, this appears as follows.
\begin{center}
(MInd){\hspace{2em}} \newSequentRule{\Gamma\vdash{}P(Z) \hspace{3em} \Gamma,\,k:\nat,\,P(k)\vdash{}P(S\,k)}{\Gamma\vdash\forall{}j\!:\!\nat.\,P(j)}
\end{center}
Read the rule as follows: To show that a property $P$ of natural numbers holds
for all natural numbers, show $P(Z)$ holds and then, assuming $P(j)$ holds for
some arbitrary $j\in\nat$ show $P(S\,j)$ holds as well. Note that this is an
instance of well-founded induction using the well-founded relation for natural
numbers, restated using the successor function in place of adding one: $i
\prec_{\nat}{}j\definedAs j = S\,i$.

How did the base case arise?  Look at the rule for well-founded induction where
the type $A$ is specialized to $\nat$ and the relation is the immediate
successor relation ($\prec_{\nat}$).
\begin{center}
\newSequentRule{\Gamma,\,j\!:\!\nat,\,\forall{}k\!:\!\nat.\; k\prec_{\nat}{}j \Rightarrow P(k) \vdash{}P(j) }{\Gamma\vdash\forall{}j\!:\!\nat.\,P(j)}
\end{center}
Since $j\in\nat$ we know $j = Z$ or $ j = S\,i$ for some $i\in\nat$. Do a case
split on $j$ giving two subgoals:
\[\begin{array}{ll}
i.) & \Gamma,Z:\nat,\forall{}k:\nat.\,k \prec_{\nat}{}Z \Rightarrow P(k) \vdash P(Z) \\
ii.) & \Gamma,i:\nat,\forall{}k:\nat.\,k \prec_{\nat}{}(S\,i) \Rightarrow P(k) \vdash P(S\,i) \\
\end{array}\]
For (i) note that the antecedent in the induction hypothesis ($k\prec_{\nat}{}Z$) is always false and so the implication is vacuously true. 
Thus $\forall{}k:\nat.\,k \prec_{\nat}{}Z \Rightarrow P(k)$ is trivially true and adds no information to our assumptions. Also, we already know $Z:\nat$ so (i) simplifies to the following:
\[\begin{array}{ll}
i.) & \Gamma\vdash P(Z) \\
\end{array}\]

For $(ii.)$ note that by the definition of $\prec_{\nat}$, if
$k\prec_{\nat}S\,i$ then $k=i$ and so $S\,i = S\,k$.  We do not need $i$ at
all, nor do we need the quantifier because the predecessor of $S\, k$ is just
$k$ itself. Using these facts we can simplify $(ii.)$ to the following:

\[\begin{array}{ll}
ii.) & \Gamma,k:\nat,P(k) \vdash P(S\,k) \\
\end{array}\]

This yields the ordinary rule for proof by mathematical induction (MInd).

\subsection{Structural Induction in General}

Structural induction is an instance of well-founded induction where the
well-founded relation on pairs of terms of type $T$, $s\prec_{T}t $ is
interpreted to mean that $s$ is an immediate subterm of $t$ ($s$ is a child of
$t$).  The immediate subterm relation is not transitive, but as noted above,
well-founded relations need not be.  It is easy to see that for finite
instances of a recursive data type this definition yields a well-founded
relation. Also note that $\prec_{\nat}$, as defined above, is the immediate
subterm relation for the type $\nat$.

We can justify the structural induction principle for a particular type $T$ by
noting that instances of a recursive type must have been generated by one of
the constructors.

For a recursive type $T$, the fact that instances of $T$ must have been
generated by one of the constructors, together with simplifications based
on the definition of $\prec_{T}$, can be used to justify the structural induction
principles we describe how to generate below.

\subsection{Generating the Induction Principle for a Recursive Type}

We build the formula expressing the induction principle for a type $T$ directly
from its data type declaration.  In Appendix A there is Haskell code which does
this.

Consider a parameterized type declaration of the following form:
\[{\mathit{data}}\; T\; [V_1,\cdots,V_k] = C_0\;[{\mathit{typeNames}}] \,\mid \;\cdots\; \mid\, C_m\;[{\mathit{typeNames}}] \]

We build the structural induction principle in steps.  
The polymorphic type
parameters ($V_i$) may denote any type. Recall that $*$ is the kind denoting
types.  Thus, the induction principle is a universally quantified formula of 
the form\footnote{Here $\Box$ denotes a hole in the formula yet to be
  defined.}.
\[\forall{}V_1:*.\,\cdots\forall{}V_k:*. \; \Box\]
Note that $T$ has arity $k$ and so $T\,V_1\,\cdots\,V_k$ is a type.  A property
of the type is a predicate over instances of the type.  This yields the
following:
\[\forall{}V_1:*.\,\cdots\forall{}V_k:*. \forall{}P:(T\; V_1\; \cdots\;V_k)\rightarrow\bool.\;\Box\]

The goal we intend to prove is that the property holds for {\em{all}} instances
of  $T$ so we can fill in the following bit:
\[\begin{array}{l}\forall{}V_1:*.\,\cdots\forall{}V_k:*. \forall{}P:(T\; V_1\; \cdots\;V_k)\rightarrow\bool.\\
{\mbox{\hspace{.25in}}}(\Box\Rightarrow\forall{}t:(T\; V_1\; \cdots\;V_k).\;P(t))
\end{array}\]

Now, since every instance of the type is of the form $C_i$ applied to the
appropriate number and types of arguments, if we can show that, no matter which
constructor was used, the property holds, then we've shown that it holds for all
instances, no matter how the instance was constructed.

Consider a constructor declaration of the form $C_i\, [T_1,\,\cdots\,,T_j]$. This
constructor has the following type:
\[C_i:T_1\rightarrow \cdots \rightarrow T_j \rightarrow (T\; V_1\;
\cdots\;V_k)\] Note that some of the $T_i$ may be instances of the type $T$
(the one being defined) itself.  These references to $T$ are the recursive
parts of the declaration and by the well-foundedness of the immediate subterm
relation, we may assume the property holds for these instances.  The clause
constructor $C_i$ of arity $j$  is defined as follows:

% \[{\cal{F}}(C_i) \definedAs \forall{}x_1\!:\!T_1.\,\cdots\,\forall{}x_j\!:\!T_j. 
% \left(\bigwedge_{\begin{array}{c}
%    i\in\{1..j\}\\
%    TyCon(T_i)=\,T
%  \end{array}}
% \hspace{-2.5em}
% P(x_i)
% \right) \Rightarrow P(C_i \,x_1\, \cdots\, x_j)\]

\[{\cal{F}}(C_i) \definedAs \forall{}x_1\!:\!T_1.\,\cdots\,\forall{}x_j\!:\!T_j. 
\left(\;\;\;\;\;\;\bigwedge\limits_{\mathclap{i\in\{1..j\}, \atop TyCon(T_i)=\,T}}P(x_i)
\right) \Rightarrow P(C_i \;x_1\; \cdots\; x_j)\]

For each type $T_i$ that is a recursive instance of $T$, we assume $P$ holds
for that instance.  Note that the constraint on inductive hypotheses is not
that $T_i= (T\, V_1\, \cdots\, V_k)$ but simply that $T$ is the type constructor 
for the type $T_i$. This allows for types where the recursive instances 
in the type declaration do not have the same arguments in every call {\em{i.e.}} see the {\it{SwapTree}} example included below.

Putting it all together we get the following structural induction principle.
\[\begin{array}{l}\forall{}V_1:*.\;\cdots\,\forall{}V_k:*.\; \forall{}P:(T\; V_1\; \cdots\;V_k)\rightarrow\bool.\\
{\mbox{\hspace{.25in}}}\left(\bigwedge_{i\in\{1..j\}}{\cal{F}}(C_i)\right)\Rightarrow\forall{}t:(T\; V_1\; \cdots\;V_k).\;P(t)
\end{array}\]

The algorithm described here is implemented by the Haskell code in Appendix
A. It is not difficult to generalize so that mutually recursive type
declarations can be handled, but we do not present that generalization in the
undergraduate course.  The function {\it{stind}} shown in Appendix A takes a
Haskell data type representing the abstract syntax of a Haskell data declaration
and returns an instance of a formula type encoding the structural induction
principle.  Within the body of {\it{stind}}, the locally defined function
{\it{mkConstructorClause}} implements the formula transformation defined above
as ${\cal{F}}$.

\section{In the classroom}

A significant motivation for teaching induction and proofs in the context of a
functional programming course is to get students thinking in a formal way about
the programs they write.  Students are encouraged to think about the putative
theorems related to the programs they write - theorems that should hold if
their programs are correct. These theorems are intended to serve as a kind of
formally stated requirements for the functions.  

As an example, assuming $\nil$ is is a right identity for append and that
append is associative:
\[\begin{array}{l}
\forall{}m:[a].\; m \append \nil = m \\
\forall{}m,n,r:[a]. (m \append n) \append r = m \append (n \append r)\\
\end{array}
\]
show that the following theorem relating reverse and append holds for finite lists:
\[
\forall{}m,n:[a].\; reverse(m \append n) = reverse\; n \append reverse\; m
\]
This theorem gives a nice characterization of list reverse in terms of append
and illustrates a pattern of contravariant behavior that can be observed in
other contexts.  Another example of this behavior is that the inverse of the
composition of relations $S$ and $R$ is the composition of the inverses of $R$
and $S$, {\em{i.e.}} $(S \circ{}R)^{-1}=R^{-1}\circ{}S^{-1}$.

As an example of the complexity of the theorems students are expected to be
able to master in the context of a two hour final exam, the following theorems
about list functions have appeared on various final exams over the last few years.
\[\begin{array}{l}
\forall{}m:[a].\; m \append \nil = m \\
\forall{}m:[a].\; length\; m = length (reverse\; m)\\
\forall{}n,m:[a]. \;length(m\, \append\; n) = (length \;m) +\, (length\; n)\\
\forall{}m,n:[a].\; length(zip\; m \; n) = min (length \;m) (length \; n)\\
\end{array}
\]
To avoid a cascade of errors, when students are asked to prove some property by
induction in the exam setting, they are provided with the induction principle
they must use together with the definitions of the functions involved and some
auxiliary theorems.  A student who fails to correctly write down an induction
principle may well know how to correctly use one.

In addition to knowing how to do proofs by induction, students in the
functional programming course are required to be able to write down the
structural induction principles for user defined types.  The program, written
for this paper, to generate induction principles has not been previously
presented in the course but will be used when the course is next offered in the
Fall 2013 semester. As Bird \cite{Bird98} and Cousineau and Mauny
\cite{Cousineau95} have noted, examples suffice to show the pattern and that is
the method that has been used in the class until now.  With the algorithm
available, students will be able to explore more examples on their own.  Class
quizzes and exams will be used to asses if students have internalized the
method or not.

Consider the following Haskell data types:
\begin{center}
\begin{program*}
\>         data Nat = Z \vertbar S Nat\\
\>         data List a = Nil \vertbar Cons a (List a)\\
\>         data Tsil a = Snoc (Tsil a) a \vertbar Lin\\ 
\>         data BTree a = Leaf a \vertbar Fork (BTree a) (BTree a)\\
\>         data SwapTree a b = Leaf \vertbar Node a (SwapTree b a) (SwapTree b a)\\
\end{program*}
\end{center}
 
\noindent{}The following are the structural induction principles output by the
Haskell code in Appendix A for the types just given.  The Haskell {\it{show}}
functions for types and formulas were specialized to produce the LaTeX output.

\[\begin{array}{l}
Nat:\\
\begin{array}[t]{l}
\forall P:Nat \rightarrow \bool. \\
\;\;\;\;((P\; Z) \wedge \forall n_1:Nat. ((P\; n_1) \Rightarrow (P (S\ n_1)))) \Rightarrow \forall n:Nat. (P\; n)\\ 
\end{array}
\vspace{.125in}\\

List\; a:\\ 
\begin{array}[t]{l}
\forall a:*.\; \forall P:(List\; a) \rightarrow \bool.\\
\;\;\;\; ((P \; Nil) \wedge\; \forall x_1:a.\; \forall l_2:(List\; a).\\
\;\;\;\;\;\;\;\;\;\;\;\;\;\;\;((P\; l_2) \Rightarrow (P\; (Cons\; x_1\; l_2)))) \\
\;\;\;\;\;\; \Rightarrow \forall l:(List\; a).\; (P \;l)
\end{array}
\vspace{.125in}\\

{\mathit{Tsil}}\; a: \\ 
\begin{array}[t]{l}
\forall a:*.\; \forall P:({\mathit{Tsil}}\; a) \rightarrow \bool.\\
\;\;\;\; (\forall x_1:a.\; \forall t_2:(Tsil\; a).\\
\;\;\;\;\;\;\;\;\;\;\;\;\;\;\; ((P\; t_2) \Rightarrow (P\; (Snoc \; t_2 \;  x_1))) \wedge (P\; Lin)) \\
\;\;\;\;\;\; \Rightarrow \forall t:({\mathit{Tsil}}\; a).\; (P \; t)\\
\end{array}
\vspace{.125in}\\

BTree \; a: \\ 
\begin{array}[t]{l}
\forall a:*.\; \forall P:(BTree \;a) \rightarrow \bool.\\
\;\;\;\;\;\;  (\forall x_1:a.\; (P\; (Leaf\; x_1)) \\
\;\;\;\;\;\;\;\; \wedge\; (\forall t_1:(BTree\; a).\; \forall t_2:(BTree\; a).\\
\;\;\;\;\;\;\;\;\;\;\;\;\;\;\; (((P\; t_1) \wedge (P t_2)) \Rightarrow (P\; (Fork\; t_1\;t_2))))) \\
\;\;\;\;\;\;\;\;\;\;\Rightarrow \forall t:(BTree\; a).\; (P\; t)\\
\end{array}
\vspace{.125in}\\

SwapTree\; a\; b:\\ 
\begin{array}[t]{l}
\forall a:*.\, \forall b:*.\, \forall P:(SwapTree\; a\; b) \rightarrow \bool.\\
\;\;\;\; ((P\; Leaf) \\
\;\;\;\;\;\; \wedge \;(\forall x_1:a.\, \forall s_2:(SwapTree\; b \;a).\, \forall s_3:(SwapTree\; b \;a).\\
\;\;\;\;\;\;\;\;\;\;\;\;\;\;\;(((P\; s_2) \wedge (P\; s_3)) \Rightarrow (P\; (Node\; x_1\;s_2\;s_3))))) \\
\;\;\;\;\;\;\;\;\;\Rightarrow \forall s:(SwapTree\; a\; b).\, (P\; s)\\
\end{array}

\end{array}
\]

\noindent{}Now consider some seemingly pathological examples;  these data types
are not recursive.
\begin{center}
\begin{program*}
\>                                 data Bool = T \vertbar F \\
\>                                 data Maybe a = Nothing \vertbar Just a\\
% \> data Rose a = Node a Nil \vertbar Node a (List (Rose a)) \\
\end{program*} 
\end{center}
\noindent{}The formulas produced by the method described above yield the
following ``induction'' principles.

\[\begin{array}{l}
Bool:\\
 \forall P:Bool \rightarrow \bool.\; ((P\; T) \wedge (P\; F)) \Rightarrow \forall b:Bool.\; (P\; b)
\vspace{.125in}\\
Maybe\; a:\\
\forall a:*.\; \forall P:(Maybe\; a) \rightarrow \bool.\\
\;\;\;\; ((P \; Nothing) \wedge \forall x_1:a.\; P (Just\; x_1)) \Rightarrow \forall m:(Maybe\; a).\; (P\; m)\\
% Rose\; a: &
% \forall a:*.\; \forall P:(Rose\; a) \rightarrow \bool.\\
% & \;\;\;\; \forall x_1:a.\; \forall x_2:(List\; (Rose\; a)).\; (P \;(Node\; x_1\;x_2)) \\
% & \;\;\;\; \;\; \Rightarrow \forall r:(Rose\; a).\; (P\; r)
\end{array}\]

The structural induction principles are generated by case analysis (on the
constructors) and by including the appropriate induction hypotheses for each
case. If there is no recursion in the type definition, the induction principle
reduces to case analysis.  The resulting formulas are theorems whether there is
recursion or not.

Induction in a lazy language like Haskell is somewhat complicated by the fact
that all types $T$ are inhabited by the undefined value $\bot$.  With regards
to well-founded relations on terms of type $T$, for all finite terms $s$,
$\bot\prec_{T}s$ and $\bot\not\prec_{T}\bot$. Case analysis on the natural
numbers yields two cases, one for numbers constructed from the constant $Z$ and
the other for numbers constructed by the successor function $S$.  To prove a
property of lazy natural numbers an additional case is added to show that
$(P\;\bot)$ holds.  This extra case arises naturally from the schema of
well-founded induction in the same way the simplified case for $Z$ does when
the case analysis splits to include the possibility of $\bot$.  The structural
induction principles can be extended to work on these {\em{pointed}} types
\cite[pp.310]{Mitchell96} simply by adding a clause $P(\bot)$ which must be
shown to hold in addition to the others.

To give the reader a sense of the difficulty, the following question appeared on
a recent final exam and was worth 12 points out of a possible 100.\\

\begin{tabular}{|ll|}
\hline 
&\ \\
1.) & [12 points] Write structural induction principles for the following Haskell data-types.{\mbox{\hspace{.15in}}} \\
&\ \\
& data STree a = Leaf $|$ Node a (STree a) (STree a)\\
&\ \\
& data Lambda c =  Var String \\
& \hspace{6.5em} $|$ Const c \\
& \hspace{6.5em}  $|$ Ap (Lambda c) (Lambda c) \\
& \hspace{6.5em}  $|$ Abs String (Lambda c)\\
&\ \\
\hline
\end{tabular}
\ \\
\goodbreak\noindent{}Students typically do well on these questions.

\section{Conclusion}

Students learning functional programming are in a unique position to be able to
prove properties about the programs they write during the development process.
It is often the case that putative properties of the functions serve as
specifications for the functions and can be used to verify their correctness.
Fundamental properties about recursive types can be verified in the context of
a functional programming course that are virtually impossible to do in the
imperative setting.  However, if they do not have the ability define structural
induction principles for newly defined types, students are left wanting in their
skills.

At the University of Wyoming we have been presenting these methods for years in
the functional programming course (COSC 3015). Students practice the methods in
homework assignments and there are often questions on the final examination
requiring them to write structural induction principles for types they have not
seen before. On the most recent exam they were required to write induction
principles for a type representing lambda terms and for a tree structure with
three kinds of nodes.

The methods described here have been widely implemented in the theorem proving
community and deserve to be better known in the functional programming
community.

\bibliographystyle{eptcs}
\bibliography{bibliography}
\newpage
\noindent{\bf{Appendix A}} Haskell code to generate structural induction principles from a Data declaration. 

\begin{program*}
% \> module SInd where \\
\> import Data.Char \\
% \> unwrap s = take (length s -2) (drop 1 s) \\
% \> wrap lbracket rbracket s = lbracket ++ s ++ rbracket \\
% \> wrapP = wrap "(" ")" \\
% \> rewrap lbracket rbracket s = wrap lbracket rbracket (unwrap s) \\
% \> rewrapP s = wrapP (unwrap s) \\
% \> unquote = filter (/='\"')  \\
% \> spaceSeparated s = " " ++ unquote (map ($\backslash${}c $\rightarrow$ if c == ',' then ' ' else c)  (unwrap s)) \\
% \>  \\
% \> showApplication s args = \\
% \>   case (rewrapP \$ show \$ map show args) of \\
% \>      "()" $\rightarrow$ s \\
% \>      a    $\rightarrow$ wrapP \$ s ++ (spaceSeparated a) \\
% \>  \\
\> data Type = Star $\mid$ Simple String [Type] $\mid$ Tuple [Type] $\mid$ Arrow Type Type \\
\>   deriving (Eq, Show)  \\
% \> instance Show Type where \\
% \>   show Star = "*" \\
% \>   show (Simple s args) = showApplication s args \\
% \>   show (Tuple types) = rewrapP \$ show (map show types) \\
% \>   show (Arrow t1 t2) = wrapP \$ show t1 ++ " $\rightarrow$ " ++ show t2 \\
% \>  \\
\> type CName = String \\
\> type TName = String \\
\> data Data = Data TName [TName] [(CName, [Type])] deriving (Eq,Show) \\
\> data Formula = FTrue \\
\>       $\mid$ Pred String [Formula] \\
\>       $\mid$ And Formula Formula \\
\>       $\mid$ Implies Formula Formula \\
\>       $\mid$ Forall String Type Formula  \\
\>    deriving (Eq,Show)  \\
% \> instance Show Formula where \\
% \>   show FTrue = "True" \\
% \>   show (Pred p args) = showApplication p args \\
% \>   show (And f1 f2) = wrapP \$ show f1 ++ " \& " ++ show f2 \\
% \>   show (Implies f1 f2) = wrapP \$ show f1 ++ " => " ++ show f2 \\
% \>   show (Forall x t f) = "All " ++ x ++ ":" ++ show t ++ ". " ++ show f \\
\> conjoin = foldr1 ($\backslash${}f fs $\rightarrow$ And f fs) \\
\> forall = foldr ($\backslash${}(v,ty) more $\rightarrow$ Forall v ty more) \\
\> mkFormulaVars = map ($\backslash${}v $\rightarrow$ Pred v [{\hspace{.125em}}])  \\
\> numberedVars vars = map ($\backslash${}(x,i) $\rightarrow$ x ++ (show i)) (zip vars [1..]) \\
\> stind (Data tname targs constructors) = \\
\>   let indVarName = map toLower (take 1 tname) in  \\
\>   let varName = "x" in  \\
\>   let newType = Simple tname (map ($\backslash${}t $\rightarrow$ Simple t [{\hspace{.125em}}]) targs) in \\
\>   let prefix body =  \\
\>           forall (Forall "P" (Arrow newType (Simple "Bool" [{\hspace{.125em}}])) body)  \\
\>                  (zip targs (repeat Star)) in \\
\>   let mkConstructorClause (c, types) =  \\
\>           if null types then  \\
\>              Pred "P" [Pred c [{\hspace{.125em}}]]  \\
\>           else  \\
\>              let arity = length types in  \\
\>              let vars = numberedVars \\
\>                    (map ($\backslash${}(Simple name \_) $\rightarrow$ \\
\>                                if name == tname then indVarName else varName) types) in  \\
\>              let varsXTypes = zip vars types in  \\
\>              let indVars = mkFormulaVars \$ \\
\>                    map fst (filter ($\backslash${}(\_, Simple t \_) $\rightarrow$ t == tname) varsXTypes) in  \\
\>              let antecedents = conjoin (map ($\backslash${}t $\rightarrow$ Pred "P" [t]) indVars) in  \\
\>              let concl =  Pred "P" [Pred c (mkFormulaVars vars)] in  \\
\>              let universal body = forall body varsXTypes  in  \\
\>                case indVars of  \\
\>                  [{\hspace{.125em}}] $\rightarrow$  universal concl  \\
\>                  \_  $\rightarrow$  universal (Implies antecedents concl) in \\
\>   let antecedent = conjoin (map mkConstructorClause constructors) in \\
\>   let concl = Forall indVarName newType (Pred "P" [Pred indVarName [{\hspace{.125em}}]]) in \\
\>       prefix (Implies antecedent concl) \\
% \>  \\
% \> -- a few example types -------------------------------------------------------------------------------------------- \\
% \>  \\
% \> treeTy = Data "Tree" ["a"] [("Leaf",[{\hspace{.125em}}]),("Node", [Simple "a" [{\hspace{.125em}}], Simple "Tree" [Simple "a" [{\hspace{.125em}}]], Simple "Tree" [Simple "a" [{\hspace{.125em}}]]])] \\
% \> tree1Ty = Data "Tree" ["a"] [("Leaf",[Simple "a" [{\hspace{.125em}}]]),("Node", [Simple "Tree" [Simple "a" [{\hspace{.125em}}]], Simple "Tree" [Simple "a" [{\hspace{.125em}}]]])] \\
% \> tree2Ty = Data "IRTree" ["a","b"] [("Leaf",[{\hspace{.125em}}]),("Node", [Simple "a" [{\hspace{.125em}}], Simple "IRTree" [Simple "b" [{\hspace{.125em}}], Simple "a" [{\hspace{.125em}}]], Simple "IRTree [Simple "a" [{\hspace{.125em}}],Simple "b" [{\hspace{.125em}}]]])] \\
% \> maybeTy = Data "Maybe" ["a"] [("Nothing",[{\hspace{.125em}}]),("Just", [Simple "a" [{\hspace{.125em}}]])] \\
% \> boolTy = Data "Bool" [{\hspace{.125em}}] [("True",[{\hspace{.125em}}]),("False", [{\hspace{.125em}}])] \\
% \> intTy = Data "N" [{\hspace{.125em}}] [("N", [Simple "Int" [{\hspace{.125em}}]])] \\
% \> listTy = Data "List" ["a"] [("Nil", [{\hspace{.125em}}]), ("Cons",[Simple "a" [{\hspace{.125em}}], Simple "List" [Simple "a" [{\hspace{.125em}}]]])] 

\end{program*}

\end{document}

%% file: header.tex
\usepackage{amssymb}
\usepackage{remark}

\newremark{theorem}{Theorem}[section]
\newremark{Rule}{Rule}[section]
 \newremark{fact}{Fact}[section]
\newremark{definition}{Definition}[section]
 \newremark{slogan}{Slogan}[section]
 \newremark{axiom}{Axiom}[section]
\newcommand{\nat}{{\mathbb{N}}}
\newcommand{\bool}{{\mathbb{B}}}

\newcommand{\newSequentRule}[2]{{\begin{tabular}{c}{${#1}$}{}\\\hline{${#2}$}{}\end{tabular}}}

\newcommand{\definedAs}{\;{\stackrel{\rm def}{=}}\;}

{\obeyspaces\global\let =\ }

\newenvironment{bogustabbing}{\begin{tabbing}\={\mbox{\hspace{10em}}}\=\=\=\kill}%
{\end{tabbing}}

\newenvironment{program*}{\sf\obeyspaces\begin{bogustabbing}}{\end{bogustabbing}}
\newenvironment{program**}{\it\obeyspaces\begin{bogustabbing}}{\end{bogustabbing}}
\newenvironment{smallprogram*}{\hspace{2.5em}\small \it\obeyspaces\begin{bogustabbing}}{\end{bogustabbing}\vspace{-.0625in}}

\newcommand{\vertbar}{$|$}

\newcommand{\append}{\,{\texttt{++}}\,}
\newcommand{\nil}{{\texttt{[{\hspace{.125em}}]}}}

%% file: TFPIE13_Caldwell_proceedings.bbl
\begin{thebibliography}{10}
\providecommand{\bibitemdeclare}[2]{}
\providecommand{\surnamestart}{}
\providecommand{\surnameend}{}
\providecommand{\urlprefix}{Available at }
\providecommand{\url}[1]{\texttt{#1}}
\providecommand{\href}[2]{\texttt{#2}}
\providecommand{\urlalt}[2]{\href{#1}{#2}}
\providecommand{\doi}[1]{doi:\urlalt{http://dx.doi.org/#1}{#1}}
\providecommand{\bibinfo}[2]{#2}

\bibitemdeclare{book}{Bertot04}
\bibitem{Bertot04}
\bibinfo{author}{Yves \surnamestart Bertot\surnameend} \&
  \bibinfo{author}{Pierre \surnamestart Cast\'eran\surnameend}
  (\bibinfo{year}{2004}): \emph{\bibinfo{title}{Interactive Theorem Proving and
  Program Development. Coq'Art: The Calculus of Inductive Constructions}}.
\newblock \bibinfo{series}{Texts in Theoretical Computer Science},
  \bibinfo{publisher}{Springer Verlag}, \doi{10.1007/978-3-662-07964-5}.

\bibitemdeclare{book}{Bird98}
\bibitem{Bird98}
\bibinfo{author}{Richard \surnamestart Bird\surnameend} (\bibinfo{year}{1998}):
  \emph{\bibinfo{title}{Introduction to Functional Programming using Haskell}},
  \bibinfo{edition}{second} edition.
\newblock \bibinfo{publisher}{Prentice-Hall}.

\bibitemdeclare{article}{Burstall69}
\bibitem{Burstall69}
\bibinfo{author}{R.~M. \surnamestart Burstall\surnameend}
  (\bibinfo{year}{1969}): \emph{\bibinfo{title}{Proving Properties of Programs
  by Structural Induction}}.
\newblock {\sl \bibinfo{journal}{The Computer Journal}}
  \bibinfo{volume}{12}(\bibinfo{number}{1}), pp. \bibinfo{pages}{41--48},
  \doi{10.1093/comjnl/12.1.41}.

\bibitemdeclare{unpublished}{Chlipala13}
\bibitem{Chlipala13}
\bibinfo{author}{Adam \surnamestart Chlipala\surnameend}:
  \emph{\bibinfo{title}{Certified Programming with Dependent Types}}.
\newblock \bibinfo{note}{Feb. 12, 2013, availabe online
  {\tt{http://adam.chlipala.net/cpdt/}}}.

\bibitemdeclare{book}{Cousineau95}
\bibitem{Cousineau95}
\bibinfo{author}{Guy \surnamestart Couineau\surnameend} \&
  \bibinfo{author}{Michel \surnamestart Mauny\surnameend}
  (\bibinfo{year}{1995}): \emph{\bibinfo{title}{The Functional Approach to
  Programming}}.
\newblock \bibinfo{publisher}{Cambridge University Press}.

\bibitemdeclare{book}{Felleisen_design}
\bibitem{Felleisen_design}
\bibinfo{author}{Matthias \surnamestart Felleisen\surnameend},
  \bibinfo{author}{Robert~Bruce \surnamestart Findler\surnameend},
  \bibinfo{author}{Matthew \surnamestart Flatt\surnameend} \&
  \bibinfo{author}{Shriram \surnamestart Krishnamurthi\surnameend}
  (\bibinfo{year}{2001}): \emph{\bibinfo{title}{How to design programs: an
  introduction to programming and computing}}.
\newblock \bibinfo{publisher}{MIT Press}, \bibinfo{address}{Cambridge, MA,
  USA}.

\bibitemdeclare{book}{Gallier85}
\bibitem{Gallier85}
\bibinfo{author}{Jean~H. \surnamestart Gallier\surnameend}
  (\bibinfo{year}{1985}): \emph{\bibinfo{title}{Logic for computer science:
  foundations of automatic theorem proving}}.
\newblock \bibinfo{publisher}{Harper \& Row Publishers, Inc.},
  \bibinfo{address}{New York, NY, USA}.

\bibitemdeclare{book}{Hudak00}
\bibitem{Hudak00}
\bibinfo{author}{Paul \surnamestart Hudak\surnameend} (\bibinfo{year}{2000}):
  \emph{\bibinfo{title}{The {H}askell School of Expression: Learning Functional
  Programming theough Multimedia}}.
\newblock \bibinfo{publisher}{Cambridge University Press}.

\bibitemdeclare{book}{Hutton07}
\bibitem{Hutton07}
\bibinfo{author}{Graham \surnamestart Hutton\surnameend}
  (\bibinfo{year}{2007}): \emph{\bibinfo{title}{Programming in {H}askell}}.
\newblock \bibinfo{publisher}{Cambridge University Press},
  \doi{10.1017/CBO9780511813672}.

\bibitemdeclare{book}{Lipovaca11}
\bibitem{Lipovaca11}
\bibinfo{author}{Miran \surnamestart Lipovaca\surnameend}
  (\bibinfo{year}{2011}): \emph{\bibinfo{title}{Learn You a {H}askell for Great
  Good: A Beginner's Guide}}.
\newblock \bibinfo{publisher}{No Starch Press}.

\bibitemdeclare{book}{Mitchell96}
\bibitem{Mitchell96}
\bibinfo{author}{John \surnamestart Mitchell\surnameend}
  (\bibinfo{year}{1996}): \emph{\bibinfo{title}{Foundations for Programming
  Langauges}}.
\newblock \bibinfo{publisher}{MIT Press}.

\bibitemdeclare{book}{Nipkow02}
\bibitem{Nipkow02}
\bibinfo{author}{Tobias \surnamestart Nipkow\surnameend},
  \bibinfo{author}{Lawrence~C. \surnamestart Paulson\surnameend} \&
  \bibinfo{author}{Markus \surnamestart Wenzel\surnameend}
  (\bibinfo{year}{2002}): \emph{\bibinfo{title}{Isabelle/HOL --- A Proof
  Assistant for Higher-Order Logic}}.
\newblock {\sl \bibinfo{series}{LNCS}} \bibinfo{volume}{2283},
  \bibinfo{publisher}{Springer}.

\bibitemdeclare{book}{paulson90}
\bibitem{paulson90}
\bibinfo{author}{Lawrence~C \surnamestart Paulson\surnameend}
  (\bibinfo{year}{1990}): \emph{\bibinfo{title}{Logic and computation:
  interactive proof with Cambridge LCF}}.
\newblock \bibinfo{publisher}{Cambridge University Press}.

\bibitemdeclare{book}{Pierce02}
\bibitem{Pierce02}
\bibinfo{author}{Benjamin \surnamestart Pierce\surnameend}
  (\bibinfo{year}{2002}): \emph{\bibinfo{title}{Types and Programming
  Langauges}}.
\newblock \bibinfo{publisher}{MIT Press}.

\bibitemdeclare{book}{Reade89}
\bibitem{Reade89}
\bibinfo{author}{Chris \surnamestart Reade\surnameend} (\bibinfo{year}{1989}):
  \emph{\bibinfo{title}{Elements of Functional Programming}}.
\newblock \bibinfo{publisher}{Addison Wesley}.

\bibitemdeclare{article}{Haskell98}
\bibitem{Haskell98}
\bibinfo{author}{et.~al. \surnamestart Simon~{Peyton Jones}\surnameend}
  (\bibinfo{year}{2003}): \emph{\bibinfo{title}{The {Haskell} 98 Language and
  Libraries: The Revised Report}}.
\newblock {\sl \bibinfo{journal}{Journal of Functional Programming}}
  \bibinfo{volume}{13}(\bibinfo{number}{1}), pp. \bibinfo{pages}{1--255}.

\bibitemdeclare{book}{Thompson99}
\bibitem{Thompson99}
\bibinfo{author}{Simon \surnamestart Thompson\surnameend}
  (\bibinfo{year}{1999}): \emph{\bibinfo{title}{Haskell: The Craft of
  Functional Programming}}, \bibinfo{edition}{second} edition.
\newblock \bibinfo{publisher}{Addison Wesley}.

\bibitemdeclare{book}{Winskel93}
\bibitem{Winskel93}
\bibinfo{author}{Glynn \surnamestart Winskel\surnameend}
  (\bibinfo{year}{1993}): \emph{\bibinfo{title}{The Formal Semantics of
  Programming Langauges: An Introduction}}.
\newblock \bibinfo{publisher}{MIT Press}.

\end{thebibliography}
